\documentclass[final,5p,times,twocolumn]{elsarticle}

\usepackage{graphicx}
\usepackage{amssymb}
\usepackage{amsmath}
\usepackage{balance}  
\usepackage{multicol}
\usepackage{tikz}
\usepackage{gensymb}
\usepackage{stmaryrd}
\usepackage{color,soul}
\usepackage{subcaption}
\usepackage{booktabs}
\usepackage[hyphens]{url}
\usepackage[hidelinks]{hyperref}

\newcommand{\fpp}{\textsc{fpp}}
\newcommand{\isep}{\textsc{isep}}

\newtheorem{defin}{Definition}

\journal{arXiv}

\begin{document}

\begin{frontmatter}



\title{Bloom filter variants for multiple sets: a comparative assessment}


\author[disi]{Luca~Calderoni\corref{cor1}}
\ead{luca.calderoni@unibo.it}

\author[disi]{Dario~Maio}
\ead{dario.maio@unibo.it}

\author[ucc]{Paolo~Palmieri}
\ead{p.palmieri@cs.ucc.ie}

\cortext[cor1]{Corresponding author}
\address[disi]{Department of Computer Science and Engineering, University of Bologna, Italy}
\address[ucc]{Department of Computer Science, University College Cork, Ireland}

\begin{abstract}
In this paper we compare two probabilistic data structures for association queries derived from the well-known Bloom filter: the shifting Bloom filter (ShBF), and the spatial Bloom filter (SBF). With respect to the original data structure, both variants add the ability to store multiple subsets in the same filter, using different strategies.
We analyse the performance of the two data structures with respect to false positive probability, and the inter-set error probability (the probability for an element in the set of being recognised as belonging to the wrong subset). As part of our analysis, we extended the functionality of the shifting Bloom filter, optimising the filter for any non-trivial number of subsets. We propose a new generalised ShBF definition with applications outside of our specific domain, and present new probability formulas.
Results of the comparison show that the ShBF provides better space efficiency, but at a significantly higher computational cost than the SBF.
\end{abstract}


\end{frontmatter}

\section{Introduction}\label{sec:intro}

The term \emph{big data} informally refers to data sets that are so large and complex that cannot be processed efficiently using traditional data structures and algorithms. The recent introduction of inexpensive information-gathering devices has fuelled the growth of pervasive sensing, which led to an increase in the size and number of data sets to be computed. Where traditional, deterministic data structures are inadequate to process big data, probabilistic data structures have been widely adopted by computer scientists as a suitable alternative \cite{DBLP:journals/comsur/TarkomaRL12}. While standard data structures are designed to store elements and answer deterministically to queries, probabilistic data structures introduce an error probability. This drawback, however, is balanced by higher space efficiency and lower computational burden, which are crucial in big data applications \cite{DBLP:journals/tocs/ChangDGHWBCFG08}.

Most probabilistic data structures, and specifically those relying on Bloom filters, use hash functions to randomize and compactly represent a set of items. The possibility of collisions introduces the potential for errors in the stored information, however the error probability is generally known and can be maintained under an arbitrary threshold by tuning the data structure parameters. Compared to errorless approaches, probabilistic data structures use significantly less memory, and have constant query time. They also usually support union and intersection operations and can therefore be easily parallelised.

Bloom filters (BF) are arguably the most prominent data structure belonging to this category. They were first introduced by Bloom \cite{DBLP:journals/cacm/Bloom70} in 1970, and were extensively used during the last decades. Bloom filters were designed to support membership queries, i.e., to determine whether a given element is a member of a given set or not. The Bloom filter always determines positively if an element is in the set, while elements outside the set are generally determined negatively, but are subject to a (bounded) false positive error probability \cite{DBLP:journals/ipl/Grandi18}.

Several modifications of Bloom filters have been proposed over the years \cite{DBLP:journals/corr/abs-1804-04777}, with the purpose of adding additional functionalities. In this paper we compare two of these variants: the shifting Bloom filter (ShBF), and the spatial Bloom filter (SBF). Both variants allow the filter to store multiple sets\footnote{We note here that the relevant literature often mixes the definition of multi-set (as in multiple sets) and multiset. In this paper, we use the term multiset following the mathematical definition of a set which permits the coexistence of multiple instances of any element. In order to remove the ambiguity with multi-set, we only use the full expression ``multiple sets'' for datasets which include elements of different sets.} rather than a single one, and enable membership queries over these sets (called \emph{association queries}). ShBF and SBF use different strategies to achieve this: the ShBF uses additional hash functions to determine a shift in the positions within the filter for each set, while the SBF writes a different index value for each set.

\subsection{Related Works}\label{sec:related}

Bloom filters and the derived data structures have been used in a number of database systems applications over the years \cite{DBLP:reference/db/2018}. Due to their ability to efficiently know whether or not a key belongs to a set of keys, they are widely adopted for first level indexes \cite{DBLP:journals/tocs/ChangDGHWBCFG08} and for query caching \cite{DBLP:journals/ton/FanCAB00}. They were extensively tested over random access memories, traditional hard drives and, more recently on solid state drives. Bloom filters were also applied to semi-structured data queries \cite{DBLP:conf/vldb/WangJLY04}, range queries and so forth.

Concerning association queries, following the classification proposed in \cite{DBLP:journals/corr/abs-1804-04777}, a number of Bloom filter variants proposed in the literature allow for multiple sets to be stored in the same filter. However, most data structures implement this functionality using a naive strategy: that is, using multiple instances of the data structure, one for each set.

Many such structures were proposed in the context of network routing strategies, where Bloom filters are widely used to address packet filtering and forwarding, and the application scenario often requires the use of multiple sets \cite{DBLP:journals/cn/GeravandA13, DBLP:journals/im/BroderM03}. For instance, the \emph{longest prefix match} solution proposed in \cite{DBLP:conf/sigcomm/DharmapurikarKT03} is one of the first strategies using the naive approach of one separate data structure for each set. In general, multiple BF strategies are viable when there is some a-priori knowledge regarding the number of sets to be mapped (and potentially the approximate number of items per set). This condition is often required for the design of hardware routing boards, where Bloom filters are parallelised and may thus be queried simultaneously. Several proposed solutions follow this principle: the IPv6 lookup scheme discussed in \cite{DBLP:conf/infocom/SongHKL09} or the packet forwarding strategy presented in \cite{DBLP:conf/infocom/ChangLF04} are widely cited examples. The difference Bloom filter (DBF) is similar in this regard as it requires additional data structures linked to the main Bloom filter \cite{DBLP:conf/icc/YangTGGYL17}. However, as the number of sets increases this approach becomes intuitively inefficient, as it increases the overhead linearly to the number of sets. It is also unsuitable when the number of sets cannot be predicted, or changes over time. We therefore exclude such Bloom filter variants from the comparison we present in this paper, to focus instead on structures that allow for multiple sets by design. A number of variants of the Bloom filter allowing multiple sets exist in the literature: shifting Bloom filters \cite{Yang:2016:SBF:2876473.2876476}, combinatorial Bloom filters \cite{DBLP:conf/infocom/HaoKLS09,DBLP:journals/ton/HaoKLS12}, spatial Bloom filters \cite{DBLP:conf/cisc/0001CM14,DBLP:journals/comcom/Calderoni0M15}, the kBF \cite{DBLP:conf/infocom/XiongYCH14}, and the Bloomier filter \cite{DBLP:conf/soda/ChazelleKRT04, DBLP:conf/esa/CharlesC08}. However, only a few are designed specifically for storing multiple sets. The kBF is instead a Bloom filter designed for key-value storage, with applications on approximate state machines; while Bloomier filters are a data structure for static support lookup tables. While both data structures can implement multiple sets, we do not include them in the comparison for reasons explained below. In the case of the kBF, the focus of the data structure is to store a value associated to each element of the originating set (following the key-value paradigm, where the key is the element and the value is the information stored in the data structure). The value is stored in a 32-bits space, divided into 29 bits used for encoding and a 3-bit counter. While we can use the value for storing a set identifier (and therefore group the elements into subsets) this is not the core functionality of the data structure, and, as it implies writing a 32-bit value for each element, is relatively inefficient. Similarly, the core functionality of Bloomier filters is not association queries on multiple sets: instead, Bloomier filters are designed for computing arbitrary functions defined only on the originating set. It is technically possible to implement a Bloomier filter defined on a membership function returning different values for elements of different subsets (and indeed this is used as an example by the authors). However, Bloomier filters can be implemented on arbitrary functions, which is a more complex problem and therefore introduce a significant overhead. In particular, Bloomier filters allow dynamic updates to the function, and also use two separate tables within the data structure, increasing memory usage. Filters that are designed for the purpose of storing multiple sets normally adopt either of two different strategies: making the filter non-binary (to allow  for the set information to be stored in a filter position) as in the case of the spatial Bloom filter; or using an increased number of hash functions, where different hashes are used for different sets. The latter strategy is used by both the shifting Bloom filters and the combinatorial Bloom filters. For the reasons detailed above, we restrict the comparative assessment presented in this paper to the data structures that are designed specifically for membership queries over multiple sets, commonly referred to as association queries. As our objective is to compare the two main strategies to do so (detailed above), we select the spatial Bloom filter to represent the non-binary filter approach, and the shifting Bloom filter, which is the most recent structure following the multiple hashes approach.

\subsection{Contribution}\label{sec:contrib}

In this paper we present a comparative assessment of two probabilistic data structures for association queries, the shifting Bloom filter and the spatial Bloom filter. In the case of the shifting Bloom filter, we also provide a new, generalised definition that allows an unlimited number of sets to be stored (contrary to the original definition that focuses on two sets only), and we provide updated error probability formulas to reflect this change. The comparative assessment is based on an experimental analysis of the behaviour of the two data structures over large test datasets. Extensive tests have been performed using comparable implementations of both primitives. The empirical results validate the theoretical analysis of the error probabilities, space efficiency and computational costs.

\section{Shifting Bloom Filter}\label{sec:shbf}

The shifting Bloom filter \cite{Yang:2016:SBF:2876473.2876476} is a variant of Bloom filters which supports membership queries as well as association queries and multiplicity queries. However, this data structure cannot provide all the aforementioned features at once. The user needs to select which kind of queries the filter should answer to, and construct it according to procedures that are specific to each functionality. The interrogation procedures are also feature-specific. As the comparison proposed in this work focuses on association queries, we limit our discussion to ShBF filters set up for this specific type of queries. In the following we refer to the ShBF built for association queries simply as ShBF. The ShBF as proposed in the original work by Yang et al. \cite{Yang:2016:SBF:2876473.2876476} focuses only on two sets, and provides a mechanism for identifying elements that are present in both sets (therefore the two sets are not disjoint, and their intersection can be $\neq \emptyset$). However, the limitation of being able to store only two sets is a direct consequence of allowing the two originating sets to be non-disjoint, and requires the use of an auxiliary data structure thus increasing memory usage. We believe that this limitation significantly reduces the potential application scenarios for the ShBF, and therefore in this paper we extend the ShBF data structure to allow for an unlimited number of originating sets. However, in doing so we remove the possibility to store sets with intersecting elements, and we allow for disjoint sets only. We feel the benefits of our proposed modification far outweight the downsides, and we note that intersections can potentially be re-introduced using a naive strategy (defining an intersection as a separate set). This modification also allows for a more meaningful comparison of the ShBF with the SBF data structure, with can also store an unlimited number of disjoint sets.

The shifting Bloom filter allows multiple sets to be stored by shifting the position at which the values are stored in the filter: that is, an \emph{offset} is added to the output of each hash function, and the offset specifies which set the element belongs to. The offset is not static, but is determined by an additional hash function, so that each set uses a different hash function to determine the offset. In practice, this means that two elements of the same set will have different offsets, but these offsets are both calculated using the same hash function, which is set-specific. For memory-efficiency reasons, Yang et al. propose in the original paper to limit the offset space (that is, the output size of this additional hash function), so that the position pointed by the hash function plus the offset will be in the same memory word as the position pointed by the hash function only (without the offset). The authors do not specify a particular size for the memory word. While this approach can reduce memory reads when only two sets are stored, in the proposed generalisation to unlimited number of sets a small word size can significantly increase the number of elements for which the set cannot be determined with accuracy.

\begin{figure*}[h!tb]
    \centering
    \begin{subfigure}[b]{0.48\textwidth}
        \scalebox{0.50}{
        		\input{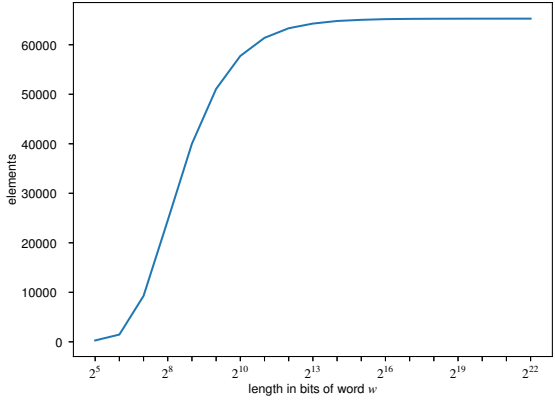}
        	}
        \caption{\small Correctly recognised elements.}
        \label{fig:w-correct}
    \end{subfigure}
    ~ 
    \begin{subfigure}[b]{0.48\textwidth}
        \scalebox{0.50}{
        		\input{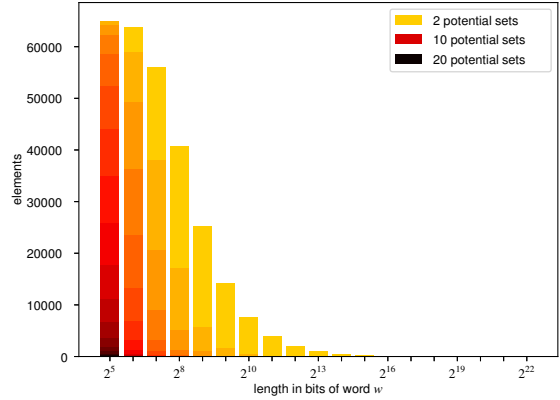}
        	}
        \caption{\small Elements returning multiple potential sets.}
        \label{fig:w-error}
    \end{subfigure}
    \caption{The word size.}\label{fig:w}
\end{figure*}

In Figure \ref{fig:w} we provide the results of an experiment where multiple word lengths are used over the same test dataset. For this experiment, we store $255$ sets each containing $256$ elements, for a total of $65280$ elements in a filter of length $2^{23}$ bits. As the two figures clearly depict, a small word value ($2^{10}$) significantly increases the number of errors returned by the filter (a formal definition of such errors is provided in the following). The best results are obtained for word sizes of $2^{16}$ or above. Considering that the original goal of reducing the memory reads cannot be achieved with such a high word size, we propose here to discard the limitation on the offset space entirely. As shown by the experiment, limiting the offset space to any meaningful word size would negatively affect the correctness of the association queries result. Moreover, removal of the word requirement reduces the computational burden by eliminating a modulo operation for each offset calculation.

In the following, we provide a formal definition of the proposed generalised ShBF, as described above. In particular, the definition we propose differs from the original ShBF as proposed by Yang et al. \cite{Yang:2016:SBF:2876473.2876476} by introducing an unlimited number of disjoint sets, and removing the offset space limitation.

\begin{defin}
Given the originating sets $\Delta_1, \Delta_2, \dots, \Delta_s$ to be represented in the filter, let $\bar{S}$ be the union set $\bar{S}=\bigcup_{\Delta_i \in S}\Delta_i$ and $S$ be the set of sets such that $S=\left\{\Delta_1, \dots, \Delta_s\right\}$. Let $H=\left\{h_1, \ldots, h_k\right\}$ be a set of $k$ hash functions such that each $h_i \in H : \left\{0, 1\right\}^* \rightarrow \left\{1, \ldots , m\right\}$, that is, each hash function takes binary strings as input and outputs a number uniformly chosen in $\left\{1, \ldots, m\right\}$. Let also $H^{\Ydown}=\left\{h^{\Ydown}_1, \ldots, h^{\Ydown}_{s-1}\right\}$ be a set of $s-1$ hash functions such that each $h^{\Ydown}_i \in H^{\Ydown} : \left\{0, 1\right\}^* \rightarrow \left\{1, \ldots , m\right\}$, where $m$ is a positive integer.

We define a \emph{shifting Bloom filter} $B^{\Ydown}\left(S\right)$ over $S$ as the set
\begin{equation}\label{eq:shbf-def}
B^{\Ydown}\left(S\right) = \bigcup_{\delta \in \bar{S}, h \in H} h(\delta)+ o(\delta) \enspace , 
\end{equation}
where the offset $o(\delta)$ is intended as follows:
\begin{equation}\label{eq:shbf-offset}
o\left(\delta\right) = \left\{
  \begin{array}{l l}
    0 \quad \quad \quad \text{if } \delta \in \Delta_1\\
    h^{\Ydown}_i(\delta) \quad \text{if } \delta \in \Delta_{i+1}, i \neq 0
  \end{array}
\right. \enspace .
\end{equation}
\end{defin}
A shifting Bloom filter $B^{\Ydown}\left(S\right)$ can be represented as a binary vector $b^{\Ydown}$ composed of $2m$ bits (or \emph{cells}), where the $i$-th bit
\begin{equation}\label{eq:shbf-bits}
b^{\Ydown}\left[i\right] = \left\{
  \begin{array}{l l}
    1 \quad \text{if } i \in B^{\Ydown}(S)\\
    0 \quad \text{if } i \not\in B^{\Ydown}(S)
  \end{array}
\right. \enspace .
\end{equation}

In the following, when referring to a shifting Bloom filter, we refer to its vector representation $b^{\Ydown}$.

The ShBF is built as follows. Initially all bits are set to $0$. Then, for each $\Delta_i \in S$, for each element $\delta \in \Delta_i$ and for each $h \in H$ we calculate $h(\delta)+o(\delta) = i$, and set the corresponding $i$-th bit of $b^{\Ydown}$ to $1$. Thus, $2m$ bits are needed in order to store $b^{\Ydown}$. As discussed in Section \ref{sec:comparison}, we implemented a circular ShBF version composed of $m$ bits.

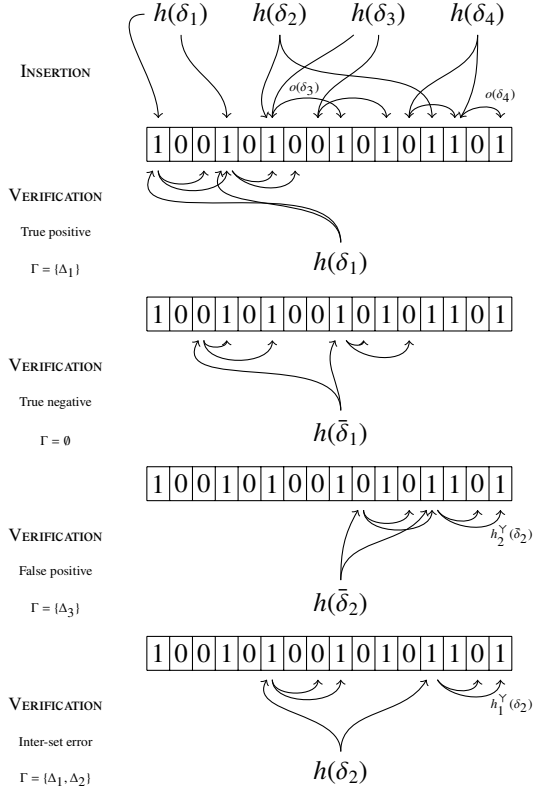
\begin{figure}[htb]
\centering
\begin{tikzpicture}[scale=0.3]



	\node[right] at (18,15.5) {\scriptsize $ \textsc{Parameters:} \qquad k=2 \qquad m=16 \qquad w=4 \qquad s=3$};
	\node[right] at (18,14) {\scriptsize $ \textsc{Hash:} \qquad\qquad\quad H=\{h_1,h_2\} \qquad H^{\Ydown}=\{h^{\Ydown}_1,h^{\Ydown}_2\}$};
	\node[right] at (18,12.5) {\scriptsize $ \textsc{Sets:} \qquad\qquad\quad \Delta_1=\{\delta_1,\delta_2\} \qquad \Delta_2=\{\delta_3\} \qquad \Delta_3=\{\delta_4\}$};
	\node[right] at (18,11) {\scriptsize $\qquad\qquad\qquad\quad S=\Delta_1 \cup \Delta_2 \cup \Delta_3 \qquad \bar{\delta}_1, \bar{\delta}_2 \not\in \bar{S} $};

	
	\node at (21,5.5) {\scriptsize \textsc{Insertion}};
	\node at (26.5,8) (anchor1) {$h(\delta_1)$};
	\node at (30.83,8) (anchor2) {$h(\delta_2)$};
	\node at (35.16,8) (anchor3) {$h(\delta_3)$};
	\node at (39.5,8) (anchor4) {$h(\delta_4)$};

	
	\draw[xstep=1.0,ystep=1.5,black,thin] (25,1.5) grid (41,3);	
	
	\node at (25.5,2.25) {$1$};
	\node at (26.5,2.25) {$0$};	
	\node at (27.5,2.25) {$0$};		
	\node at (28.5,2.25) {$1$};		
	\node at (29.5,2.25) {$0$};		
	\node at (30.5,2.25) {$1$};		
	\node at (31.5,2.25) {$0$};		
	\node at (32.5,2.25) {$0$};
	\node at (33.5,2.25) {$1$};		
	\node at (34.5,2.25) {$0$};
	\node at (35.5,2.25) {$1$};
	\node at (36.5,2.25) {$0$};
	\node at (37.5,2.25) {$1$};
	\node at (38.5,2.25) {$1$};
	\node at (39.5,2.25) {$0$};
	\node at (40.5,2.25) {$1$};	
	
	\node at (25.5,3) (anchor1-1) {};
	\node at (28.5,3) (anchor1-4) {};
	\draw (anchor1) edge[out=180,in=90,->] (anchor1-1);
	\draw (anchor1) edge[out=270,in=90,->] (anchor1-4);
	
	\node at (30.5,3) (anchor2-6) {};
	\node at (37.5,3) (anchor2-13) {};
	\draw (anchor2) edge[out=270,in=120,->] (anchor2-6);
	\draw (anchor2) edge[out=270,in=90,->] (anchor2-13);
	
	\node at (30.5,3) (anchor3-6) {};
	\node at (32.5,3) (anchor3-8) {};
	\node at (33.5,3) (anchor3-9) {};
	\node at (35.5,3) (anchor3-11) {};
	\draw (anchor3) edge[out=220,in=90,->] (anchor3-6);
	\draw (anchor3-6) edge[out=90,in=90,->] (anchor3-9);
	\node at (31.9,4.7) {\tiny $o(\delta_3)$};
	\draw (anchor3) edge[out=270,in=90,->] (anchor3-8);
	\draw (anchor3-8) edge[out=90,in=90,->] (anchor3-11);
	
	\node at (36.5,3) (anchor4-12) {};
	\node at (38.5,3) (anchor4-14) {};
	\node at (40.5,3) (anchor4-16) {};
	\draw (anchor4) edge[out=270,in=90,->] (anchor4-12);
	\draw (anchor4-12) edge[out=90,in=90,->] (anchor4-14);
	\draw (anchor4) edge[out=270,in=60,->] (anchor4-14);
	\draw (anchor4-14) edge[out=60,in=90,->] (anchor4-16);
	\node at (40.5,4.4) {\tiny $o(\delta_4)$};
	
	
	\node at (21,0) (ver1) {\scriptsize \textsc{Verification}};
	\node [below of = ver1, node distance = 0.5cm] (ver1-1) {\tiny True positive};
	\node [below of = ver1-1, node distance = 0.5cm] () {\tiny $\Gamma=\{\Delta_1\}$};
	\node at (33.5,-3) (anchor5) {$h(\delta_1)$};
	
	\node at (25.5,1.5) (anchor5-1) {};
	\node at (27.5,1.5) (anchor5-3) {};
	\node at (28.5,1.5) (anchor5-4) {};
	\node at (30.5,1.5) (anchor5-6) {};
	\node at (31.5,1.5) (anchor5-7) {};
	\draw (anchor5) edge[out=90,in=240,->] (anchor5-1);
	\draw (anchor5-1) edge[out=270,in=270,->] (anchor5-3);
	\draw (anchor5-1) edge[out=270,in=270,->] (anchor5-4);
	\draw (anchor5) edge[out=90,in=240,->] (anchor5-4);
	\draw (anchor5-4) edge[out=300,in=270,->] (anchor5-6);
	\draw (anchor5-4) edge[out=300,in=270,->] (anchor5-7);

	
	\draw[xstep=1.0,ystep=1.5,black,thin] (25,-6) grid (41,-4.5);	
	
	\node at (25.5,-5.25) {$1$};
	\node at (26.5,-5.25) {$0$};	
	\node at (27.5,-5.25) {$0$};		
	\node at (28.5,-5.25) {$1$};		
	\node at (29.5,-5.25) {$0$};		
	\node at (30.5,-5.25) {$1$};		
	\node at (31.5,-5.25) {$0$};		
	\node at (32.5,-5.25) {$0$};
	\node at (33.5,-5.25) {$1$};		
	\node at (34.5,-5.25) {$0$};
	\node at (35.5,-5.25) {$1$};
	\node at (36.5,-5.25) {$0$};
	\node at (37.5,-5.25) {$1$};
	\node at (38.5,-5.25) {$1$};
	\node at (39.5,-5.25) {$0$};
	\node at (40.5,-5.25) {$1$};

	\node at (21,-7.5) (ver2) {\scriptsize \textsc{Verification}};
	\node [below of = ver2, node distance = 0.5cm] (ver2-1) {\tiny True negative};
	\node [below of = ver2-1, node distance = 0.5cm] () {\tiny $\Gamma=\emptyset$};
	\node at (33.5,-10.5) (anchor6) {$h(\bar{\delta}_1)$};
	
	\node at (27.5,-6) (anchor6-3) {};
	\node at (33.5,-6) (anchor6-9) {};
	\node at (28.5,-6) (anchor6-4) {};
	\node at (30.5,-6) (anchor6-6) {};
	\node at (34.5,-6) (anchor6-10) {};
	\node at (36.5,-6) (anchor6-12) {};
	\draw (anchor6) edge[out=90,in=240,->] (anchor6-3);
	\draw (anchor6-3) edge[out=270,in=270,->] (anchor6-4);
	\draw (anchor6-3) edge[out=270,in=270,->] (anchor6-6);
	\draw (anchor6) edge[out=90,in=240,->] (anchor6-9);
	\draw (anchor6-9) edge[out=300,in=270,->] (anchor6-10);
	\draw (anchor6-9) edge[out=300,in=270,->] (anchor6-12);

	
	\draw[xstep=1.0,ystep=1.5,black,thin] (25,-13.5) grid (41,-12);
	
	\node at (25.5,-12.75) {$1$};
	\node at (26.5,-12.75) {$0$};
	\node at (27.5,-12.75) {$0$};
	\node at (28.5,-12.75) {$1$};
	\node at (29.5,-12.75) {$0$};
	\node at (30.5,-12.75) {$1$};
	\node at (31.5,-12.75) {$0$};
	\node at (32.5,-12.75) {$0$};
	\node at (33.5,-12.75) {$1$};
	\node at (34.5,-12.75) {$0$};
	\node at (35.5,-12.75) {$1$};
	\node at (36.5,-12.75) {$0$};
	\node at (37.5,-12.75) {$1$};
	\node at (38.5,-12.75) {$1$};
	\node at (39.5,-12.75) {$0$};
	\node at (40.5,-12.75) {$1$};

	\node at (21,-15) (ver3) {\scriptsize \textsc{Verification}};
	\node [below of = ver3, node distance = 0.5cm] (ver3-1) {\tiny False positive};
	\node [below of = ver3-1, node distance = 0.5cm] () {\tiny $\Gamma=\{\Delta_3\}$};
	\node at (33.5,-18) (anchor7) {$h(\bar{\delta}_2)$};
	\node at (41,-15) {\tiny $h^{\Ydown}_2(\bar{\delta}_2)$};
	
	\node at (34.5,-13.5) (anchor7-10) {};
	\node at (36.5,-13.5) (anchor7-12) {};
	\node at (37.5,-13.5) (anchor7-13) {};
	\node at (39.5,-13.5) (anchor7-15) {};
	\node at (40.5,-13.5) (anchor7-16) {};
	\draw (anchor7) edge[out=90,in=240,->] (anchor7-10);
	\draw (anchor7-10) edge[out=270,in=270,->] (anchor7-12);
	\draw (anchor7-10) edge[out=270,in=270,->] (anchor7-13);
	\draw (anchor7) edge[out=90,in=240,->] (anchor7-13);
	\draw (anchor7-13) edge[out=300,in=270,->] (anchor7-15);
	\draw (anchor7-13) edge[out=300,in=270,->] (anchor7-16);

	
	\draw[xstep=1.0,ystep=1.5,black,thin] (25,-21) grid (41,-19.5);
	
	\node at (25.5,-20.25) {$1$};
	\node at (26.5,-20.25) {$0$};
	\node at (27.5,-20.25) {$0$};
	\node at (28.5,-20.25) {$1$};
	\node at (29.5,-20.25) {$0$};
	\node at (30.5,-20.25) {$1$};
	\node at (31.5,-20.25) {$0$};
	\node at (32.5,-20.25) {$0$};
	\node at (33.5,-20.25) {$1$};
	\node at (34.5,-20.25) {$0$};
	\node at (35.5,-20.25) {$1$};
	\node at (36.5,-20.25) {$0$};
	\node at (37.5,-20.25) {$1$};
	\node at (38.5,-20.25) {$1$};
	\node at (39.5,-20.25) {$0$};
	\node at (40.5,-20.25) {$1$};

	\node at (21,-22.5) (ver4) {\scriptsize \textsc{Verification}};
	\node [below of = ver4, node distance = 0.5cm] (ver4-1) {\tiny Inter-set error};
	\node [below of = ver4-1, node distance = 0.5cm] () {\tiny $\Gamma=\{\Delta_1,\Delta_2\}$};
	\node at (33.5,-25.5) (anchor8) {$h(\delta_2)$};
	\node at (41,-22.5) {\tiny $h^{\Ydown}_1(\delta_2)$};
	
	\node at (30.5,-21) (anchor8-6) {};
	\node at (32.5,-21) (anchor8-8) {};
	\node at (33.5,-21) (anchor8-9) {};
	\node at (37.5,-21) (anchor8-13) {};
	\node at (39.5,-21) (anchor8-15) {};
	\node at (40.5,-21) (anchor8-16) {};
	\draw (anchor8) edge[out=90,in=240,->] (anchor8-6);
	\draw (anchor8-6) edge[out=270,in=270,->] (anchor8-8);
	\draw (anchor8-6) edge[out=270,in=270,->] (anchor8-9);
	\draw (anchor8) edge[out=90,in=240,->] (anchor8-13);
	\draw (anchor8-13) edge[out=300,in=270,->] (anchor8-15);
	\draw (anchor8-13) edge[out=300,in=270,->] (anchor8-16);
	
\end{tikzpicture}

\caption{Insertion of three originating sets $\Delta_1, \Delta_2$ and $\Delta_3$ and verification of elements in a shifting Bloom filter of length $m=16$, featuring two hash functions ($k=2$). With respect to the verification process, it is possible to observe true positives, true negatives, false positives and inter-set errors. Each scenario is depicted in sequence.}\label{fig:ShBFexample}
\end{figure}

The verification procedure is formalized as follows. Let us suppose to test an element $\delta \in \mathcal{E}$ against the filter (where $\mathcal{E}$ represents a generic domain), in order to learn which originating set it belongs to (or whether it does not belong to any originating set). First of all we need to compute $k$ hash digests, i.e., $\forall h \in H$, we compute $h(\delta)$. Then, as each originating set is combined with a specific hash offset, we need to check separately whether the element belongs to each set. We check whether $\delta \in \Delta_i$ if

\begin{equation}\label{eq:a-in-Si}
\forall h \in H, b^{\Ydown}\left[h(\delta)+o(\delta)\right] = 1 \enspace .
\end{equation}

During the verification process of the set $\Delta_i$, should we find an index $i$ such that $b^{\Ydown}[i]=0$, we can conclude $\delta \notin \Delta_i$ and we proceed to another set $\Delta_j \in S$.
Conversely, when the condition (\ref{eq:a-in-Si}) is fulfilled, we add $\Delta_i$ to the set of positives matches $\Gamma$.
At the end of the verification procedure three are the possible scenarios:

\begin{itemize}
\item $\Gamma = \emptyset$. Then we may assert $\delta \notin \bar{S}$.
\item $|\Gamma| = 1, \Delta_i \in \Gamma$. Hence $\delta \in \Delta_i$.
\item $|\Gamma| > 1$. Hence $\delta$ belongs to one among the sets included in $\Gamma$.
\end{itemize}

Both the second and the third case are subject to a false positive probability, as ShBF preserves the probabilistic nature of the classic Bloom filter. Concerning the latter case, please also note that it is not possible to assign a different weight to those sets included in $\Gamma$. Consequently, the element $\delta$ has the same probability to belong to each set, resulting in an inter-set error.

A schematization of the insertion and verification procedures of an ShBF is proposed in Figure \ref{fig:ShBFexample}.

\subsection{False positives}\label{sub:shbf-fp}
As we redefined the ShBF data structure it is important to investigate its behaviour in terms of false positives.

\begin{defin}
Given a filter $b^{\Ydown}$, a false positive event occurs when the verification procedure performed on an element $\delta \not \in \bar{S}$ terminates with a non-empty set of positives matches, i.e. $\Gamma \neq \emptyset$.
\end{defin}

As we may see, a false positive event reported by a ShBF may pertain several originating sets, due to the verification procedure which performs a separate lookup for each originating set. Consequently, we define a false positive event on a specific set as follows:

\begin{defin}
Given a filter $b^{\Ydown}$, a false positive event on a specific originating set $\Delta_i$ occurs when the verification procedure performed on an element $\delta \not \in \bar{S}$ terminates with the set $\Delta_i$ included in the set of positives matches, i.e. $\Delta_i \in \Gamma$.
\end{defin}

Since during the verification procedure each set is checked for membership through a separate filter lookup performed on the same data structure, each originating set has the same probability of being included in the set of positives matches $\Gamma$. Specifically, the set-specific false positive probability ($\fpp_i^{\Ydown}$) in a ShBF coincides with the overall false positives probability of a common BF, assuming both filters were filled with the same amount of items, i.e. $b^{\Ydown}$ was filled with $n=|\bar{S}|$ items:

\begin{equation}\label{eq:shbf-fpp-i}
\fpp_i^{\Ydown} = \left(1 - \left(1 - \frac{1}{m}\right)^{kn}\right)^k \enspace .
\end{equation}

Conversely, the overall false positives probability ($\fpp^{\Ydown}$) is increased due to the multiple filter lookups performed during the verification procedure. In order for the verification procedure to produce a false positive, it sufficient that one among the $s$ set-specific lookups produces a set-specific false positive. As the probability for this event to occur is known ($\fpp_i^{\Ydown}$), we may address this problem using the well known success-failure scheme (Bernoulli trials). Given an event which occurs with probability $p$, the probability to observe $b$ successes among $a$ trials is:

\begin{equation}
{a \choose b} p^b (1-p)^{a-b} \enspace .
\end{equation}

The probability to report at least one set-specific false positive (i.e. at least one success) among $s$ trials is the probability of the certain event minus the probability to never report a set-specific false positive among $s$ trials. Hence

\begin{equation}\label{eq:shbf-fpp-1}
\fpp^{\Ydown} = 1 - \left( {s \choose 0} {\fpp_i^{\Ydown}}^0 (1-\fpp_i^{\Ydown})^{s-0} \right) \enspace ;
\end{equation}

it follows that:

\begin{equation}\label{eq:shbf-fpp-2}
\fpp^{\Ydown} = 1 - \left( 1 - \left(1 - \left(1 - \frac{1}{m}\right)^{kn}\right)^k \right)^s \enspace .
\end{equation}

\subsection{Inter-set errors}\label{sub:shbf-ise}
As discussed above, when the verification procedure produces a set of positives matches composed of more than one set, it is not possible to discern which among those sets is the one that effectively contains the element. Thus, supposing $|\Gamma| = u$, we have only $1/u$ chances to assign the element to the correct set.

\begin{defin}
Given a filter $b^{\Ydown}$, an inter-set error event occurs when the verification procedure performed on an element $\delta \in \bar{S}$ terminates with a set of positives matches such that $|\Gamma|>1$.
\end{defin}

As per the classic BF, the ShBF is not affected by false negatives. Thus, when we test the filter for an element $\delta \in \Delta_i$, the correct set ($\Delta_i$) is for sure included in $\Gamma$. However, the verification procedure is composed of $s$ ShBF-lookups, one for each set. During the remaining $s-1$ lookups, should the procedure report a false-positive, the corresponding set would be wrongly added to the set of positives matches. Hence, the probability to observe an inter-set error ($\isep^{\Ydown}$) may be derived as the probability of the certain event minus the probability to never report a false positive throughout $s-1$ trials.

Following the same principle described in (\ref{eq:shbf-fpp-1}) we may derive:

\begin{equation}\label{eq:shbf-isep-1}
\isep^{\Ydown} = 1 - \left( {s-1 \choose 0} {\fpp_i^{\Ydown}}^0 (1-\fpp_i^{\Ydown})^{s-1-0} \right) \enspace ;
\end{equation}

hence

\begin{equation}\label{eq:shbf-isep-2}
\isep^{\Ydown} = 1 - \left( 1 - \left(1 - \left(1 - \frac{1}{m}\right)^{kn}\right)^k \right)^{s-1} \enspace .
\end{equation}

As the amount of correct information produced by the verification procedure depends on the number of sets included in $\Gamma$, it is also useful to derive the probability $\isep_{u_i}^{\Ydown}$ for it to produce a set of positives matches with a specific cardinality $i$. Following the aforementioned principle, it is straightforward that the probability to obtain $|\Gamma|=i$ is the probability to report exactly $i-1$ false positives among $s-1$ trials. Hence:

\begin{equation}
\isep_{u_i}^{\Ydown} = {s-1 \choose i-1} {\fpp_i^{\Ydown}}^{i-1} (1-\fpp_i^{\Ydown})^{s-i} \enspace .
\end{equation}\label{eq:shbf-isep-3}

Finally, for ease of comparison with SBF, we discuss the probability for a specific set to be involved in a inter-set error event ($\isep_i^{\Ydown}$).

\begin{defin}
Given a filter $b^{\Ydown}$, an inter-set error event on a specific originating set $\Delta_i$ occurs when the verification procedure performed on an element $\delta \in \bar{S}$ terminates with a set of positives matches such that $\Delta_i \in \Gamma$, $|\Gamma|>1$.
\end{defin}

Following the same principle discussed in Section \ref{sub:shbf-fp} concerning false positive events on specific sets, it is straightforward to note that $\isep_i^{\Ydown}$ coincides with $\fpp_i^{\Ydown}$ (see (\ref{eq:shbf-fpp-i}) for reference).

\section{Spatial Bloom Filter}\label{sec:sbf}

The spatial Bloom filter was originally introduced in \cite{DBLP:conf/cisc/0001CM14,DBLP:journals/comcom/Calderoni0M15} with the purpose of efficiently storing an arbitrary number of disjoint sets representing geographic areas. Although its first application was in the location privacy domain (as the name suggests), the data structure can store any type of element and thus represents an efficient solution for any kind of scenarios where association queries over an unlimited number of sets are required \cite{DBLP:conf/cf/0001CM17}.

Similarly to the generalised ShBF we propose earlier in this paper, SBF natively supports multiple sets and element mapping is performed through a single data structure, without any need of auxiliary data structures or meta data.

The spatial Bloom filter (SBF) is defined as follows \cite{DBLP:journals/comcom/Calderoni0M15}:
\begin{defin}\label{def:sbf}
Given the originating sets $\Delta_1, \Delta_2,\ldots, \Delta_s$ to be represented in the filter, let $\bar{S}$ be the union set $\bar{S} = \bigcup_{\Delta_i \in S} \Delta_i$ and $S$ be the set of sets such that $S = \left\{\Delta_1, \dots, \Delta_s\right\}$. Let $O$ be the strict total order over $S$ for which $\Delta_i < \Delta_j$ for $i<j$. Let also $H=\left\{h_1, \ldots, h_k\right\}$ be a set of $k$ hash functions such that each $h_i \in H : \left\{0, 1\right\}^* \rightarrow \left\{1, \ldots , m\right\}$, that is, each hash function in $H$ takes binary strings as input and outputs a random number uniformly chosen in $\left\{1, \ldots, m\right\}$. We define the \emph{spatial Bloom filter} over $\left(S,O\right)$ as the set of couples
\begin{equation}\label{eq:SB-def}
B^\#\left(S,O\right) = \bigcup_{i \in I} \enspace \langle i,\max L_i \rangle
\end{equation}
where $I$ is the set of all values output by hash functions in $H$ for elements of $\bar{S}$
\begin{equation}\label{eq:SB-i}
\quad I = \bigcup_{\delta \in \bar{S}, h \in H} h\left(\delta\right)
\end{equation}
and $L_i$ is the set of labels $l$ such that:
\begin{equation}\label{eq:SB-Li}
L_i = \left\{ l\ |\ \exists \delta \in \Delta_l, \exists h \in H : h(\delta) = i\right\} \enspace .
\end{equation}
\end{defin}

A spatial Bloom filter $B^\#\left(S,O\right)$ can be represented as a vector $b^\#$ composed of $m$ values (or \emph{cells}), where the $i$-th value
\begin{equation}\label{eq:sb-elements}
b^\#\left[i\right] = \left\{
  \begin{array}{l l}
    l \quad \text{if } \langle i,l \rangle \in B^\#\left(S,O\right)\\
    0 \quad \text{if } \langle i,l \rangle \not\in B^\#\left(S,O\right)
  \end{array}
\right. \enspace \text{\cite{DBLP:journals/comcom/Calderoni0M15}}.
\end{equation}

Throughout this paper, we refer to spatial Bloom filters using the vector representation $b^\#$.

\begin{figure}[htb]
\centering
\begin{tikzpicture}[scale=0.3]



	\node[right] at (18,15.5) {\scriptsize $ \textsc{Parameters:} \qquad k=2 \qquad m=16 \qquad s=3$};
	\node[right] at (18,14) {\scriptsize $ \textsc{Hash:} \qquad\qquad\quad H=\{h_1,h_2\}$};
	\node[right] at (18,12.5) {\scriptsize $ \textsc{Sets:} \qquad\qquad\quad \Delta_1=\{\delta_1,\delta_2\} \qquad \Delta_2=\{\delta_3\} \qquad \Delta_3=\{\delta_4\}$};
	\node[right] at (18,11) {\scriptsize $\qquad\qquad\qquad\quad S = \Delta_1 \cup \Delta_2 \cup \Delta_3 \qquad \bar{\delta}_1, \bar{\delta}_2 \not\in \bar{S} $};

	
	\node at (21,5.5) {\scriptsize \textsc{Insertion}};
	\node at (26.5,8) (anchor1) {$h(\delta_1)$};
	\node at (30.83,8) (anchor2) {$h(\delta_2)$};
	\node at (35.16,8) (anchor3) {$h(\delta_3)$};
	\node at (39.5,8) (anchor4) {$h(\delta_4)$};

	
	\draw[xstep=1.0,ystep=1.5,black,thin] (25,1.5) grid (41,3);	
	
	\node at (25.5,2.25) {$1$};
	\node at (26.5,2.25) {$0$};
	\node at (27.5,2.25) {$0$};
	\node at (28.5,2.25) {$0$};		
	\node at (29.5,2.25) {$0$};		
	\node [opacity=0.2] at (30.5,2.45) {$1$};
	\node at (30.5,2.05) {$2$};
	\node at (31.5,2.25) {$0$};
	\node at (32.5,2.25) {$2$};
	\node at (33.5,2.25) {$0$};
	\node at (34.5,2.25) {$1$};
	\node at (35.5,2.25) {$0$};
	\node at (36.5,2.25) {$3$};
	\node [opacity=0.2] at (37.5,2.45) {$1$};
	\node at (37.5,2.05) {$3$};
	\node at (38.5,2.25) {$0$};
	\node at (39.5,2.25) {$0$};
	\node at (40.5,2.25) {$0$};
	
	\node at (25.5,3) (anchor1-1) {};
	\node at (34.5,3) (anchor1-10) {};
	\draw (anchor1) edge[out=180,in=90,->] (anchor1-1);
	\draw (anchor1) edge[out=270,in=90,->] (anchor1-10);
	
	\node at (30.5,3) (anchor2-6) {};
	\node at (37.5,3) (anchor2-13) {};
	\draw (anchor2) edge[out=270,in=120,->] (anchor2-6);
	\draw (anchor2) edge[out=270,in=90,->] (anchor2-13);
	
	\node at (30.5,3) (anchor3-6) {};
	\node at (32.5,3) (anchor3-8) {};
	\draw (anchor3) edge[out=220,in=90,->] (anchor3-6);
	\draw (anchor3) edge[out=270,in=90,->] (anchor3-8);
	
	\node at (36.5,3) (anchor4-12) {};
	\node at (37.5,3) (anchor4-13) {};
	\draw (anchor4) edge[out=270,in=90,->] (anchor4-12);
	\draw (anchor4) edge[out=270,in=60,->] (anchor4-13);
	
	
	\node at (21,0) (ver1) {\scriptsize \textsc{Verification}};
	\node [below of = ver1, node distance = 0.5cm] (ver1-1) {\tiny True positive};
	\node [below of = ver1-1, node distance = 0.5cm] () {\tiny $v^\#(\delta_1) = 1$};
	\node at (33.5,-3) (anchor5) {$h(\delta_1)$};
	
	\node at (25.5,1.5) (anchor5-1) {};
	\node at (34.5,1.5) (anchor5-10) {};
	\draw (anchor5) edge[out=90,in=270,->] (anchor5-1);
	\draw (anchor5) edge[out=90,in=270,->] (anchor5-10);

	
	\draw[xstep=1.0,ystep=1.5,black,thin] (25,-6) grid (41,-4.5);	
	
	\node at (25.5,-5.25) {$1$};
	\node at (26.5,-5.25) {$0$};	
	\node at (27.5,-5.25) {$0$};		
	\node at (28.5,-5.25) {$0$};		
	\node at (29.5,-5.25) {$0$};		
	\node at (30.5,-5.25) {$2$};		
	\node at (31.5,-5.25) {$0$};		
	\node at (32.5,-5.25) {$2$};
	\node at (33.5,-5.25) {$0$};		
	\node at (34.5,-5.25) {$1$};
	\node at (35.5,-5.25) {$0$};
	\node at (36.5,-5.25) {$3$};
	\node at (37.5,-5.25) {$3$};
	\node at (38.5,-5.25) {$0$};
	\node at (39.5,-5.25) {$0$};
	\node at (40.5,-5.25) {$0$};

	\node at (21,-7.5) (ver2) {\scriptsize \textsc{Verification}};
	\node [below of = ver2, node distance = 0.5cm] (ver2-1) {\tiny True negative};
	\node [below of = ver2-1, node distance = 0.5cm] () {\tiny $v^\#(\bar{\delta}_1) = 0$};
	\node at (33.5,-10.5) (anchor6) {$h(\bar{\delta}_1)$};
	
	\node at (27.5,-6) (anchor6-3) {};
	\node at (34.5,-6) (anchor6-10) {};
	
	\draw (anchor6) edge[out=90,in=270,->] (anchor6-3);
	\draw (anchor6) edge[out=90,in=270,->] (anchor6-10);

	
	\draw[xstep=1.0,ystep=1.5,black,thin] (25,-13.5) grid (41,-12);
	
	\node at (25.5,-12.75) {$1$};
	\node at (26.5,-12.75) {$0$};
	\node at (27.5,-12.75) {$0$};
	\node at (28.5,-12.75) {$0$};
	\node at (29.5,-12.75) {$0$};
	\node at (30.5,-12.75) {$2$};
	\node at (31.5,-12.75) {$0$};
	\node at (32.5,-12.75) {$2$};
	\node at (33.5,-12.75) {$0$};
	\node at (34.5,-12.75) {$1$};
	\node at (35.5,-12.75) {$0$};
	\node at (36.5,-12.75) {$3$};
	\node at (37.5,-12.75) {$3$};
	\node at (38.5,-12.75) {$0$};
	\node at (39.5,-12.75) {$0$};
	\node at (40.5,-12.75) {$0$};

	\node at (21,-15) (ver3) {\scriptsize \textsc{Verification}};
	\node [below of = ver3, node distance = 0.5cm] (ver3-1) {\tiny False positive};
	\node [below of = ver3-1, node distance = 0.5cm] () {\tiny $v^\#(\bar{\delta}_2) = 1$};
	\node at (33.5,-18) (anchor7) {$h(\bar{\delta}_2)$};
	
	\node at (25.5,-13.5) (anchor7-1) {};
	\node at (37.5,-13.5) (anchor7-13) {};
	\draw (anchor7) edge[out=90,in=270,->] (anchor7-1);
	\draw (anchor7) edge[out=90,in=270,->] (anchor7-13);

	
	\draw[xstep=1.0,ystep=1.5,black,thin] (25,-21) grid (41,-19.5);
	
	\node at (25.5,-20.25) {$1$};
	\node at (26.5,-20.25) {$0$};
	\node at (27.5,-20.25) {$0$};
	\node at (28.5,-20.25) {$0$};
	\node at (29.5,-20.25) {$0$};
	\node at (30.5,-20.25) {$2$};
	\node at (31.5,-20.25) {$0$};
	\node at (32.5,-20.25) {$2$};
	\node at (33.5,-20.25) {$0$};
	\node at (34.5,-20.25) {$1$};
	\node at (35.5,-20.25) {$0$};
	\node at (36.5,-20.25) {$3$};
	\node at (37.5,-20.25) {$3$};
	\node at (38.5,-20.25) {$0$};
	\node at (39.5,-20.25) {$0$};
	\node at (40.5,-20.25) {$0$};

	\node at (21,-22.5) (ver4) {\scriptsize \textsc{Verification}};
	\node [below of = ver4, node distance = 0.5cm] (ver4-1) {\tiny Inter-set error};
	\node [below of = ver4-1, node distance = 0.5cm] () {\tiny $v^\#(\delta_2) = 2$};
	\node at (33.5,-25.5) (anchor8) {$h(\delta_2)$};
	
	\node at (30.5,-21) (anchor8-6) {};
	\node at (37.5,-21) (anchor8-13) {};

	\draw (anchor8) edge[out=90,in=270,->] (anchor8-6);
	\draw (anchor8) edge[out=90,in=270,->] (anchor8-13);
	
\end{tikzpicture}

\caption{Insertion of three originating sets $\Delta_1, \Delta_2$ and $\Delta_3$ and verification of elements in a spatial Bloom filter of length $m=16$, featuring two hash functions ($k=2$). With respect to the verification process, it is possible to observe true positives, true negatives, false positives and inter-set errors. Each scenario is depicted in sequence.}\label{fig:SBFexample}
\end{figure}
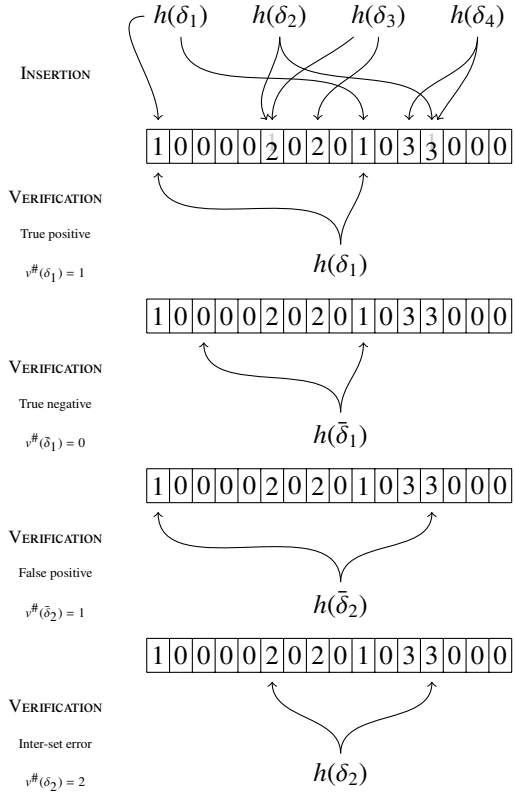

The construction procedure of a SBF may be summarized as follows \cite{DBLP:journals/comcom/Calderoni0M15}. The first step consists in setting each value in $b^\#$ to $0$. Then we compute $h\left(\delta\right) = i$ for each item $\delta \in \Delta_1$ and for each $h \in H$. The corresponding cell of the filter ($b^\#\left[i\right]$) is set to $1$ (as $1$ represents the label of $\Delta_1$). The same procedure is performed for elements included in $\Delta_2$ (writing the value $2$), then for those included in $\Delta_3$ and so forth. The construction procedure terminates when the last set $\Delta_s$ has been processed. It is important to point out that, as the construction procedure follows the strict total order defined over $S$, sets with lower labels are more likely to be overwritten than sets with higher label values, in case of collision.

Concerning collisions, we note that they occur when a specific hash function produces the same digest for two different elements (this is indeed what we commonly refer to as a hash collision), but also when two separate hash functions produce the same digest for a single element or for two distinct elements. When one among the aforementioned conditions is met, the same filter cell is written twice. We refer to this phenomenon as \emph{cell overwrite}, or simply as \emph{collision}, indifferently. An in-depth analysis concerning collisions and how they affect the probabilistic properties of this data structure is beyond the scope of this work and may be found in \cite{8271998}.

During the filter construction, the filter crosses several \emph{states} \cite{8271998}:

\begin{defin}\label{def:sbfstate}
Let us consider the spatial Bloom filter $b^{\#}$. Given $i \in L$, we say the filter is in \emph{state} $i$ (and we refer to its vector representation as $b_{\dashv i}^{\#}$) if and only if all the elements of set $\Delta_i$ have been inserted into the filter.
\end{defin}
Therefore, state $0$ ($b_{\dashv 0}^{\#}$) represents the empty filter (with all of its cells set to zero). At the end of the construction process, the filter is in state $s$ ($b_{\dashv s}^{\#}$).

In order to know whether or not an element belongs to one of the originating sets, we need to perform a single filter lookup. Given $\delta \in \mathcal{E}$ (where $\mathcal{E}$ represents a generic domain), we compute $k$ hash digest, i.e., $\forall h \in H$, we compute $h(\delta)$. We check whether $\delta \in \Delta_i$ if

\begin{equation}\label{eq:d-in-Delta-i}
\exists h \in H : b^\#\left[h(\delta)\right] = i \quad \text{and} \quad \forall h \in H, b^\#\left[h(\delta)\right] \geq i \enspace .
\end{equation}

Should one or more cells contain the value $0$, we can conclude $\delta \notin \bar{S}$, i.e., it does not belong to any of the originating sets.

The verification procedure which classifies an element $\delta \in \mathcal{E}$ (where $\mathcal{E}$ represents a generic domain) as belonging to one of the originating sets may be formalized using a functional notation as follows:

\begin{equation}
\begin{split}
v^\#: \quad & \mathcal{E} \rightarrow L_0 \\
& \delta \mapsto v^\#(\delta)
\end{split}
\end{equation}

where $L_0 = \left\lbrace 0, \dots, s \right\rbrace$.
This function processes a generic element of the domain $\mathcal{E}$ and outputs a value included in $\{0,\dots,s\}$. This integer indicates the set to which the element is supposed to belong to (if the output is $> 0$), or indicates that the element does not belong to any of the originating sets (when the output is $0$). As thoroughly investigated in \cite{8271998}, false negatives are not possible, while positive matches are subject to false-positives and inter-set errors. An exemplification of insertion and verification procedures concerning a SBF is proposed in Figure \ref{fig:SBFexample}.

\subsection{False positives}\label{sub:sbf-fp}
While an in-depth investigation concerning SBF false positives falls outside the scope of this work, it is meaningful to recall some outcome for ease of comparison with the ShBF.
Specifically, the overall false positive probability in a SBF coincides with the one of a common BF, assuming both filters were filled with the same amount of items, i.e. $|\bar{S}|=n$.
In this case, the probability to observe a false positive when we query a SBF for membership of an element which does not belong to any of the originating sets is
\begin{equation}\label{eq:sbf-fpp}
\fpp^\# = \left(1 - \left(1 - \frac{1}{m}\right)^{kn}\right)^k \enspace .
\end{equation}
As widely discussed in \cite{8271998}, this probability may be divided in a set-specific one. Specifically, the probability to report a false positive on a given set $\Delta_i$ is:
\begin{equation}
\fpp_i^\# = \left(1 - \left(1 - \frac{1}{m}\right)^{k \sum_{j=i}^s n_j}\right)^k - \sum_{j=i+1}^s \fpp_j^\# \enspace .
\end{equation}

\subsection{Inter-set errors}\label{sub:sbf-ise}
As stated above, an in-depth investigation concerning SBF inter-set errors falls outside the scope of this work. However, we report the main outcomes discussed in \cite{8271998} for ease of comparison with the ShBF.
The probability to observe a set-specific inter-set error is:

\begin{equation}\label{eq:sbf-isep-i}
\isep_i^\# = \left(1 - \left(1 - \frac{1}{m}\right)^{k n_i^{\textsc{FILL}}}\right)^k \enspace ,
\end{equation}

where $n_i^{\textsc{FILL}}$ represents the number of elements left for insertion after the filter $b^\#$ reaches the state $i$ (see \cite{8271998} for details).

Assuming $|\Delta_i| = n_i$, the overall inter-set error probability ($\isep^\#$) may be derived as the weighted sum of each set-specific probability:

\begin{equation}\label{eq:sbf-isep}
\isep^\# = \frac{1}{n} \sum_{i=1}^s n_i \isep_i^\# \enspace .
\end{equation}

\section{Comparison}\label{sec:comparison}

We compare the performances of the shifting Bloom filters and the spatial Bloom filters over the two most meaningful filter characteristics: the false positive probability and the inter-set error rate. Because of the different strategies used by the data structures to store multiple sets, these probabilities need to be evaluated with respect to two different parameters: the filter length in bits $l$ (which expresses the memory usage of the corresponding filter), and the number of cells $m$ (where each cell is a unique position within the filter). In the case of the ShBF, these two parameters coincide and $l=m$, as each cell can store either a $0$ or a $1$, following the binary nature of the original Bloom filter. It is important to point out that, by definition, ShBF is subject to an offset function which may cause indexed cells to exceed $m$. For ease of comparison, we opted for a circular implementation of the ShBF: during both construction and verification phase, cell indexing is subject to a $m$ modulus, that is, given an element $\delta$ to be inserted or tested against the filter, each hash function is evaluated in combination with the proper offset as $(h(\delta) + o(\delta))\% m$. The SBF allows instead multi-bit cells, where the data structure stores the index of the relevant set. The length of the filter in bits is therefore the number of cells times the size in bits of each cell, which is determined as the smallest power of 2 which is equal or greater than the number of sets to be stored plus one. So, if we need to store $255$ sets, $l = 8 \cdot m$, as $2^{8} = 256$.

\begin{table}[hb]
  \centering
  \caption{Test datasets used for the experimental comparison of the two datastructures.}
  \label{tab:dataset}
    \begin{tabular}{ l r r r}
      \toprule
      {dataset} & sets ($s$) & elem. per set & elements ($n$)\\
      \midrule
      uniform & 255 & 256 & 65280 \\
      random & 255  & $\left[209,298\right]$ & 65280 \\
      non-elements & - & - & 500000 \\
    \bottomrule
    \end{tabular}
\end{table}

\begin{figure*}[ht]
    \centering
    \scalebox{0.7}{
	\input{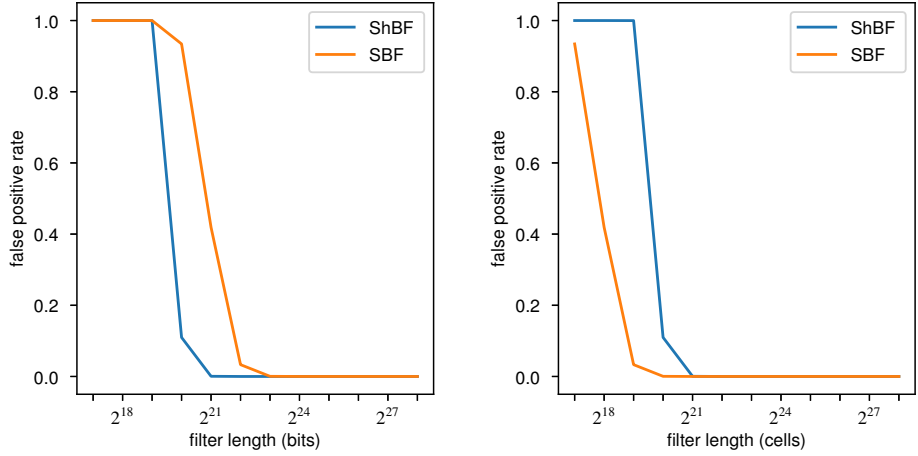}
    }
    \caption{\small False positive probability for the two data structures over a uniformly distributed dataset, calculated over the same filter length in bits (left) and the same number of cells (right).}\label{fig:fpr}
\end{figure*}

\begin{figure*}[h!]
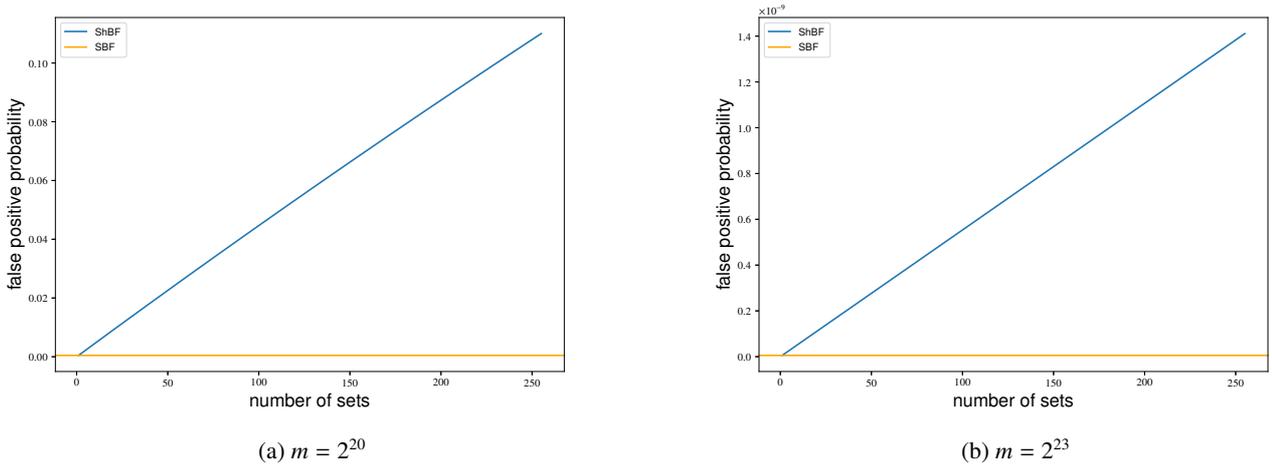

    \centering
    \begin{subfigure}[b]{0.49\textwidth}
        \scalebox{0.40}{
        		\input{test/fig_fpp_per_query_20.pgf}
        	}
        \caption{\small $m=2^{20}$}
        \label{fig:fpp-single-query-20}
    \end{subfigure}
    ~ 
    \begin{subfigure}[b]{0.49\textwidth}
        \scalebox{0.40}{
        		\input{test/fig_fpp_per_query_23.pgf}
        	}
        \caption{\small $m=2^{23}$}
        \label{fig:fpp-single-query-23}
    \end{subfigure}
    \caption{A plot of the false positive probability as expressed in (\ref{eq:shbf-fpp-2}) and (\ref{eq:sbf-fpp}) as the number of originating sets $s$ increases. The considered filter length is $m=2^{20}$ (a) and $m=2^{23}$ (b).}
    \label{fig:fpp-single-query}
\end{figure*}

We performed extensive experimental tests over implementations of both the shifting Bloom filter and the spatial Bloom filter using the test datasets described in Table \ref{tab:dataset}. Both datasets have a total of $65280$ elements distributed over $255$ sets. In the uniform dataset, each set contains exactly $256$ elements, while in the random dataset the elements are distributed randomly between sets. In order to test the false positive probability, we use an additional dataset of $500000$ ``non-elements'' that are not part of either test datasets, and therefore should ideally be detected as not belonging to any of the originating sets.

Results of the false positive probability experiment are depicted in Figure \ref{fig:fpr}. The figure plots the observed ratio of false positives over the uniform dataset. On the left hand side, the ratio is calculated over a varying filter legth in bits $l$, while on the right hand side this is done over the number of cells $m$. Given that $l=m$ for the ShBF, as described above, the experimentally observed probability remains the same, while the ratio changes for the SBF. From the experiment, we can see that the SBF performs better if we consider the number of cells $m$, resulting in a negligible number of false positives for $m \geq 2^{20}$. However, if we consider the memory space used by the filter and therefore the filter length in bits, ShBF can provide optimal results starting at $l = 2^{21}$ while the SBF at $l = 2^{23}$.

While the experiment depicted in Figure \ref{fig:fpr} was carried out keeping the number of originating sets fixed, it is important to point out that, concerning false positive probability, the SBF outperforms the ShBF as the number $s$ of originating sets increases. This condition is evident from (\ref{eq:shbf-fpp-2}) and (\ref{eq:sbf-fpp}), and is outlined in Figure \ref{fig:fpp-single-query}. In particular, for $m=20$ the false positive probability for the ShBF increases from close to $0$ for a single set to over $0.1$ for $250$ sets (Figure \ref{fig:fpp-single-query-20}). In the case of a larger filter with $m=23$, the false positive probability for the ShBF ranges from close to $0$ to $1.4 \times 10^{9}$. The false positive probability of the SBF is instead independent of the number of sets, as expected.

\begin{figure*}[thb]
    \centering
    \begin{subfigure}[b]{0.49\textwidth}
        \scalebox{0.50}{
        		\input{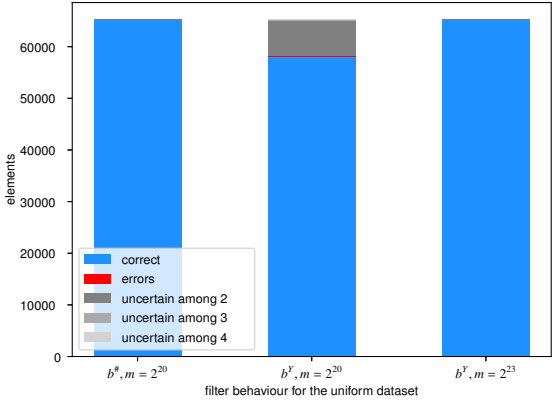}
        	}
        \caption{\small Uniform dataset.}
        \label{fig:interset-unif}
    \end{subfigure}
    ~ 
    \begin{subfigure}[b]{0.49\textwidth}
        \scalebox{0.50}{
        		\input{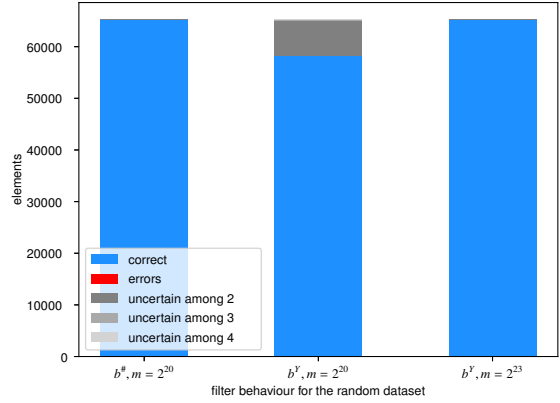}
        	}
        \caption{\small Random dataset.}
        \label{fig:interset-rand}
    \end{subfigure}
    \caption{Filter behaviour of the two data structures over two datasets, (a) with a uniform distribution of elements among the 255 sets, (b) with a random distribution. The SBF is tested over a number of cells $m=2^{20}$ and filter length in bits $l=2^{23}$ (left). The ShBF is tested over the same number of cells $m=l=2^{20}$ (centre); and over the same filter length in bits with $m=l=2^{23}$ (right). For a ShBF, the number of cells and the filter length in bits coincide.}
    \label{fig:interset}
\end{figure*}

Figure \ref{fig:interset} depicts the behaviour of the ShBF and SBF over both the uniform and random datasets with respect to inter-set errors. The bar on the left hand side for both graphs shows results for the SBF over a number of cells $m=2^{20}$ and filter length in bits $l=2^{23}$. The ShBF is tested over the same number of cells $m=2^{20}$ as the SBF in the bar at the centre; and over the same filter length in bits with $l=2^{23}$ in the bar on the right hand side. Since some outcomes concerning errors are not easily visible, results of the experiment are also provided in Table \ref{tab:ise}.

\begin{table}[h!]
\centering
\caption{Results of the experiment also depicted in Figure \ref{fig:interset}. $c$ is the number of elements whose set was correctly identified. For the SBF, the number of elements recognised as belonging to a wrong set is $e$. For the ShBF, $u_i$ is the number of elements that were identified as belonging to one out of $i$ sets. The resulting entropy $ent$ is also provided for both data structures, calculated as described in Section \ref{sec:comparison}.}
\label{tab:ise}
\footnotesize
\begin{tabular}{@{}lrrrrrrrl@{}}
\toprule
\multicolumn{1}{l}{filter} & \multicolumn{1}{c}{$m$} & \multicolumn{1}{c}{$c$} & \multicolumn{1}{l}{$e$} & \multicolumn{1}{c}{$u_2$} & \multicolumn{1}{c}{$u_3$} & \multicolumn{1}{c}{$u_4$} & \multicolumn{1}{c}{$u_5$} & \multicolumn{1}{c}{$ent$} \\ \midrule
          & \multicolumn{8}{c}{\textsc{uniform dataset}}                                                                                                                                                 \\
SBF    & 20     & 65276                 & 4                     & -                       & -                       & -                       & -                       & 0.99994     \\
ShBF   & 20     & 58174                 & -                     & 6739                    & 352                     & 15                      & 0                       & 0.94462     \\
ShBF   & 23     & 65276                 & -                     & 4                       & 0                       & 0                       & 0                       & 0.99997     \\
          & \multicolumn{8}{c}{\textsc{random dataset}}                                                                                                                                                  \\
SBF    & 20     & 65277                 & 3                     & -                       & -                       & -                       & -                       & 0.99995     \\
ShBF   & 20     & 58282                 & -                     & 6600                    & 379                     & 18                      & 1                       & 0.94536     \\
ShBF   & 23     & 65278                 & -                     & 2                       & 0                       & 0                       & 0                       & 0.99998     \\
\bottomrule
\end{tabular}
\end{table}

Given that the computational cost (in terms of number of hash computations) increases significantly as the number of sets increases for the ShBF (as discussed in the following section, and Table \ref{tab:compcost}), we performed the tests over large but compact datasets of $65280$ elements and $255$ sets. Larger datasets would not provide significantly different results from the error probability perspective, given the filter parameters $m$ and $k$ are adjusted accordingly, but would incur in a largely increased computation time. Again, we notice here that the ShBF outperforms the SBF in terms of space efficiency, but not in terms of cell numbers. It is important to note here the different behaviour of the two data structures with respect to the occurrence of an inter-set error. In the case of the ShBF, an inter-set error happens when the originating set cannot be determined exactly, because more than one potential set is returned by the filter over a query. However, the correct set is always included in the returned list. In the case of the SBF, instead, and inter-set error returns a single set, which is not the correct originating set. In order to compare the different behaviours, we introduce an \emph{entropy} metric here (modeled around the entropy concept first proposed by Shannon) which assesses the amount of information returned by a query. For the purpose of this work, we consider a correct result as having entropy $1$, an incorrect result (possible only for the SBF) as having entropy $0$, and a doubtful result (when multiple sets are returned by querying an ShBF) as $1/u$ where $u$ is the number of sets returned. Average entropy results ($ent$) for the experiment are provided in Table \ref{tab:ise}, where we call an event where $u=i$ as $u_i$. The results indicate that the ShBF has a marginally higher entropy for the same filter length $l$ for both distributions. Finally, we note here that while inter-set errors are distributed uniformly in the case of the ShBF, the SBF has a higher occurrence of inter-set errors for sets that have low index (that is, those that are entered first into the filter during construction). Therefore, the filter can be tuned to have different error probabilities for different sets, which may be an advantage in certain application scenarios.

\subsection{Computational cost}\label{sec:compcost}

In analysing the error probability of the two data structures, we referred to the memory usage of the filters. However, the operation of both data structures also implies a computational cost, which is different between the two constructions. In particular, in the following we analyse the computational cost associated with a single query to a filter. In general, cryptographic operations are the most computationally expensive, and therefore we analyse here the number of hash computations required for a query. Results are shown in Table \ref{tab:compcost}.

\begin{table}
\centering
\caption{Computational cost (expressed as number of hash computations required) for each lookup and query.}
\label{tab:compcost}
    \begin{tabular}{ l c c c }
      \toprule
      filter & lookups/ & hashes/ &  cells read / query \\
             & query    & query  & [min, max] \\
      \midrule
      ShBF & $s$ & $k+s-1$ & $\left[s, \left(s \cdot k\right)\right]$ \\
      SBF  & 1 & $k$ & $\left[1, k\right]$ \\
    \bottomrule
    \end{tabular}
\end{table}

\begin{figure}[thb]
    \centering
    \scalebox{0.55}{
	\input{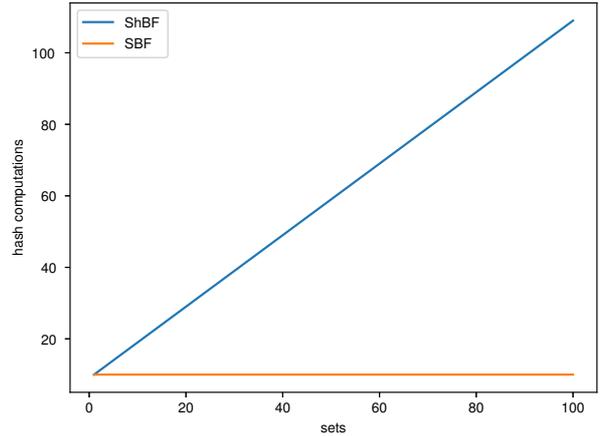}
    }
    \caption{\small Number of hash computations required for a single query over the ShBF (linear) and SBF (constant) over the number of sets.}
    \label{fig:hashes_query}
\end{figure}

A shifting Bloom filter requires $\left(k+s-1\right)$ hash digests to be computed for each query (where $k$ is the number of hash functions chosen at filter construction, and $s$ is the number of sets), and each query also involves $s$ lookups. Moreover, a single query involves reading a minimum of $s$ cells (if a 0 is returned in the first position of all set offsets) and a maximum of $\left(s \cdot k\right)$ (in the worst case scenario for which a 1 is returned for each cell read). A spatial Bloom filter, instead, requires a constant number of hash computations $k$ and a single lookup for each query, with a minimum of 1 and a maximum of $k$ cells read. We can therefore conclude that the computational cost of a ShBF is significantly higher per query than the cost of an equivalent SBF, as evident in Figure \ref{fig:hashes_query}.

\section{Conclusions}\label{sec:conclusions}
In this paper we compared the shifting Bloom filters and the spatial Bloom filters, two probabilistic data structures designed to support association queries. For the former data structure, we also provide a novel generalised definition allowing an unlimited amount of originating sets, and we discuss the resulting probabilistic model in detail. We implemented and tested both data structures over several datasets. Results show that the two data structures provide different benefits, and an adoption choice should depend on the application context. In particular, the probabilistic model of the spatial Bloom filter shows that it outperforms the shifting Bloom filter concerning both false positives and inter-set errors when the number of cells is considered. However, when effective memory consumption is considered, the shifting Bloom filter can achieve similar error rates to the spatial Bloom filter using less memory. With regards to computational costs, SBF requires a constant number of hash calculations, while ShBF requires a number of hash digest computations that increases linearly with the number of sets.

\section{Acknowledgements}
The authors would like to thank Herakliusz Oskar Lipiec, who worked on a python implementation of the shifting Bloom filter data structure as part of his final year project at University College Cork.

\bibliographystyle{elsarticle-num}
\bibliography{biblio}

\end{document}